\documentclass[journal]{IEEEtran}

\ifCLASSINFOpdf
   \usepackage[pdftex]{graphicx}
\else
   \usepackage[dvips]{graphicx}
\fi

\usepackage{booktabs}
\usepackage{caption}
\usepackage{multirow}
\usepackage{pdflscape}
\usepackage[table, svgnames, dvipsnames]{xcolor}
\usepackage{makecell}
\usepackage{cellspace}
\usepackage{tabularx,tabulary}
\usepackage{amssymb}% http://ctan.org/pkg/amssymb
\usepackage{pifont}% http://ctan.org/pkg/pifont
\usepackage{hhline}
\usepackage{pgf}
\usepackage[colorinlistoftodos]{todonotes}
\usepackage[bookmarks=false]{hyperref}
\usepackage[edges]{forest}
\usepackage{adjustbox}
\usepackage{mathtools}
\usepackage{amsmath}
\usepackage{todonotes}
\usepackage{longtable}

\newcolumntype{L}{>{\raggedright\arraybackslash}X}
\newcolumntype{C}{>{\centering\arraybackslash}X}
\newcolumntype{P}{>{\raggedright\arraybackslash}m{1.7cm}}
\hypersetup{nolinks=true}

\begin{document}

\title{V2X Misbehavior in Maneuver Sharing and Coordination Service: Considerations for Standardization}

\author{\IEEEauthorblockN{Jean-Philippe Monteuuis, Jonathan Petit, Mohammad Raashid Ansari, Cong Chen, Seung Yang,}\\
\IEEEauthorblockA{Qualcomm Technologies, Inc. \\
Boxborough, MA, USA \\
\{jmonteuu,petit,seunyang,ransari,congchen\}@qti.qualcomm.com}
}

\maketitle

\begin{abstract}
Connected and Automated Vehicles (CAV) use sensors and wireless communication to improve road safety and efficiency. However, attackers may target Vehicle-to-Everything (V2X) communication. Indeed, an attacker may send authenticated-but-wrong data to send false location information, alert incorrect events, or report a bogus object endangering other CAVs' safety. Standardization Development Organizations (SDO) are currently working on developing security standards against such attacks. Unfortunately, current standardization efforts do not include misbehavior specifications for advanced V2X services such as Maneuver Sharing and Coordination Service (MSCS). This work assesses the security of MSC Messages (MSCM) and proposes inputs for consideration in existing standards.
\end{abstract}

\IEEEpeerreviewmaketitle

\section{Introduction}
\label{sec:introduction}
Thanks to Vehicle-to-Everything (V2X) communication, road safety can be significantly improved. It enables V2X-equipped vehicles to exchange their telematics information to create awareness, especially in non-line-of-sight (NLoS) conditions. Cooperative awareness is achieved by broadcasting a message called Basic Safety Message (BSM) or Cooperative Awareness Message (CAM). Both messages contain similar information (location and kinematic state of the sender) but are defined by two different standards. BSM is defined in the Society of Automotive Engineers (SAE) J2735 standard~\cite{SAEJ2735} and CAM is defined in the European Telecommunications Standards Institute (ETSI) European Standard (EN) 302 637-2 standard~\cite{ETSIEN302637-2}\footnote{Henceforth, we will refer to both BSM and CAM services as only BSM service.}. However, BSMs do not provide the maneuver intent from the transmitting CAV. Therefore, a Maneuver Sharing and Coordination Service (MSCS) has been created to share maneuvers among V2X-enabled vehicles. Vehicles participating in the MSCS generate and consume Maneuver Sharing and Coordination Messages (MSCM) that are designed to complement the BSM/CAM service. 

BSMs and MSCMs are intended to be used to make driving decisions by an operator or an automated driving system. Due to this reason, these services become safety-critical. Thus, it is paramount that the information passed through these services are accurate. An attacker like the one discussed in Section~\ref{sec:attacker-model} can send incorrect data to affect receivers' telematics awareness negatively. Commonly, attacks on BSMs jeopardize V2X applications~\cite{monteuuis2018my}. Therefore, deploying a Misbehavior Detection System (MBDS) is mandatory to detect and protect against such attackers~\cite{defcon28}. However, very little research exists on the security of the MSCS. This paper summarizes the results of a threat assessment (TA) on the MSCS defined in SAE J3186~\cite{SAEJ3186}. Lastly, we discuss the gaps in MSCS and MBDS standards and propose items for consideration.

The structure of the paper is as follows. Section~\ref{sec:related-work} presents the standardization and academic efforts in the domain of V2X MSC and its security. Section~\ref{sec:system-model} details the system model and the MSCM. Section~\ref{sec:attacker-model} describes the attacker model considered in our TA presented in Section~\ref{sec:threat-assessment}. Section~\ref{sec:discussion} discusses standardization and research's open challenges to achieve a secure MSCS. Finally, Section~\ref{sec:conclusion} concludes this paper.

%%%%%%%%%%%%%%%%%%%%%%%%%% Related Work %%%%%%%%%%%%%%%%%%%%%%%%%
\section{Related Work}
\label{sec:related-work}
This section provides an overview of functional and security standards for MSCM. Additionally, this section includes related academic work.

\subsection{Standardization}
This section briefly introduces existing and ongoing standards from a functional and security perspective.
\begin{figure}
     \centering
     \includegraphics[width=\columnwidth]{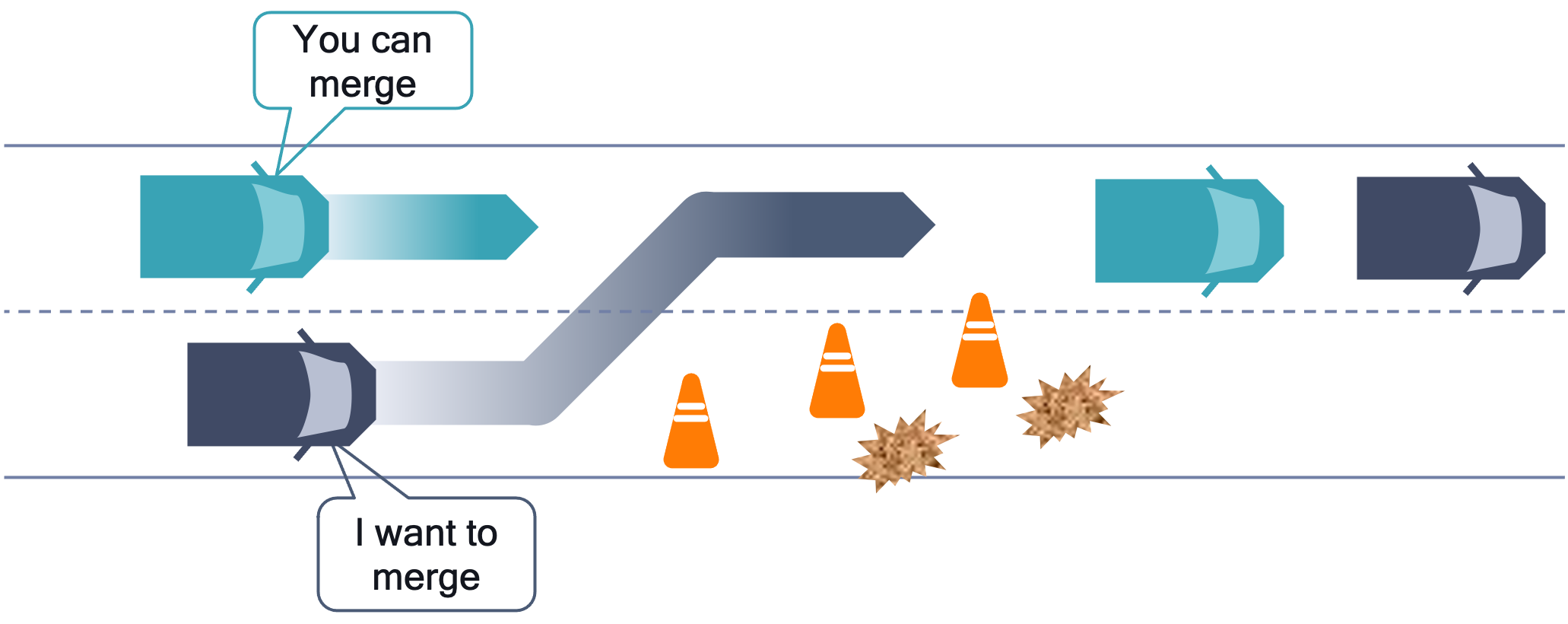}
     \caption{Use Case for Maneuver Sharing and Coordination \cite{car2019guidance}}
     \label{fig:mscm-overview}
\end{figure}

\subsubsection{Functional Standards}
The notion of MSC has been introduced in the V2X community to share maneuver intent among CAVs and smart infrastructures. As a result, each CAV can enhance its driving tasks by considering the maneuver intent of neighboring CAVs. Figure~\ref{fig:mscm-overview} illustrates a MSC scenario. This service prevents the need for a CAV to guess or predict the maneuver intent of other CAVs. Currently, several ongoing standardization initiatives exist in:
\begin{itemize}
    \item North America (SAE J3186~\cite{SAEJ3186})
    \item Europe (ETSI TR 103 578~\cite{ETSI103578} \& TS 103 561~\cite{ETSI103561})
    \item China (CSAE 157~\cite{CSAE157})
\end{itemize}
Even though the standards above have the same purpose, each document has its own set of specifications. Moreover, thus, each document might have slightly different cybersecurity threats. In this work, we present our TA of SAE J3186.

\subsubsection{Security Standard}
ETSI TS 103 759~\cite{ETSI103759} is a standard under development that defines V2X MBD and reporting activities for CAM and Decentralized Environmental Messages (DENM). The supporting TR 103 460~\cite{ETSI103460} briefly mentioned the detection and reporting of MSCM, but the details are out-of-scope of version 1 of the ETSI TS 103 759.

\begin{table}[tb!]
\centering
\caption{Status of MBD specification per message type}
\begin{tabulary}{\columnwidth}{|c|l|}
\hline
\textbf{V2X Message}
    & \multicolumn{1}{|c|}{\textbf{Misbehavior Detectors Status}}  \\
\hline   
BSM / CAM
	& Specified \\
\hline    
DENM
	& Specified\\
\hline     
MSCM
	& Specification is missing\\
\hline     
\end{tabulary}
\label{tab: ETSI Scope}
\end{table}

\subsection{Academic work}
\label{sec:rw-academic}
As far as we know, there is no research on MSCS security. However, the security of the planning stack (trajectory prediction) in autonomous driving is at an early research stage~\cite{man2022evaluating, jiao2022semi}, and, as explained in Section~\ref{sec:system-model}, MSCS and planning are related. 

In~\cite{man2022evaluating}, researchers fooled the planning stack by attacking the perception system. The considered attack scenario was as follows. The perception system of an automated vehicle incorrectly classifies an adversarial vehicle as a pedestrian due to an adversarial patch attack~\cite{chen2018shapeshifter}. As a result, the planning stack loaded the prediction model used for pedestrian trajectory instead of the one used for vehicle trajectory. The outcome was erratic movements from the car's victim and increased safety risk for the victim and its surrounding vehicles.

To our knowledge, the prior art has not studied attacks targeting cooperative planning. Thus, our work aims to fill this gap.

%%%%%%%%%%%%%%%%%%%%%%%%%% SYSTEM MODEL %%%%%%%%%%%%%%%%%%%%%%%%%
\section{System Model}
\label{sec:system-model}
This section provides an overview of the MSC system, service, and messages. Lastly, this section describes the role of V2X applications and security in the context of MSC. 

\subsection{Maneuver Sharing and Coordination System} 
\begin{figure}[t!]
    \centering
    \includegraphics[width=\columnwidth]{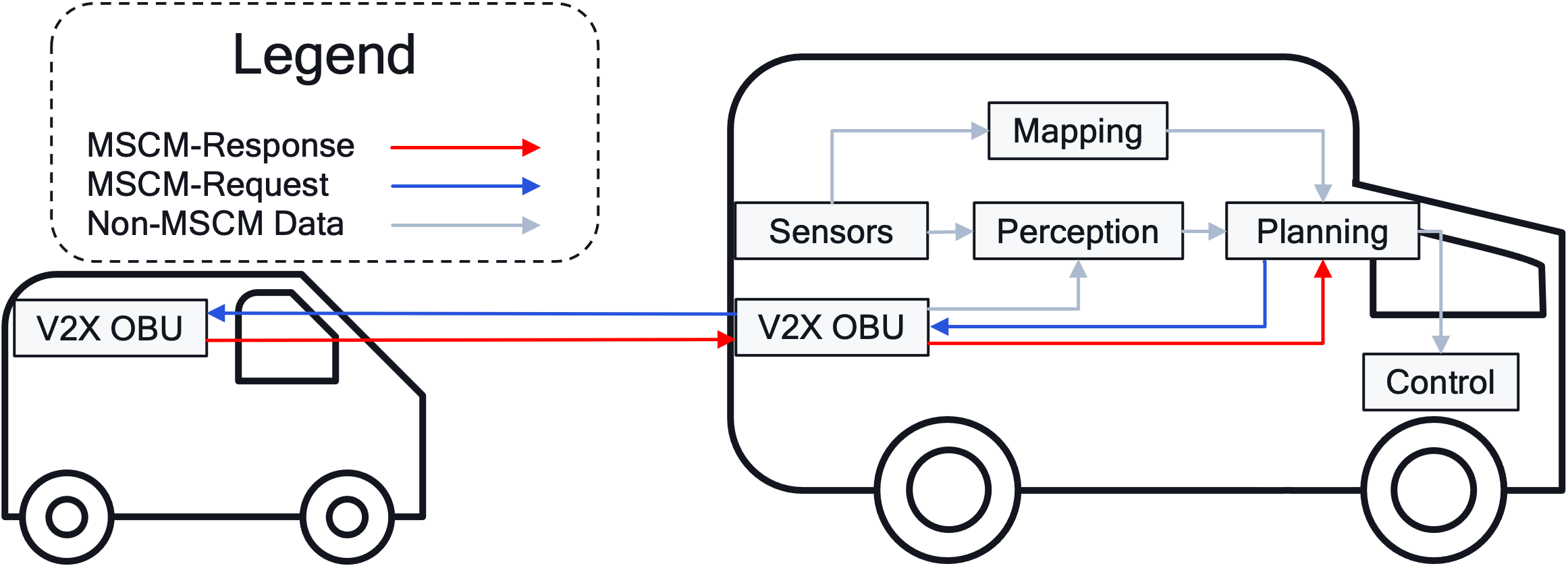}
    \caption{Maneuver Sharing and Coordination System}
    \label{fig:MSCS}
\end{figure}

As seen in Figure~\ref{fig:MSCS}, the MSC service requires a MCS system composed of a V2X On-Board Unit (OBU), sensors (e.g., camera), a mapping stack, a perception stack, a planning stack, and a control system (e.g., steering, brakes, engine). The perception stack provides a view of the ego-vehicle (depicted as a truck) and its surroundings (e.g., other vehicles). The mapping stack contains map data (e.g., roads, road lanes, and buildings). Lastly, the planning stack decides and updates the vehicle's maneuvers. The planning stack relies on several components. First, the perception stack provides road obstacles to be avoided by the planning stack. Then, the mapping stack provides all the areas where the vehicle can navigate. Lastly, the V2X OBU provides the maneuver intent (MSCM) from surrounding CAVs and transmits the maneuver intent from the planning stack to the surrounding vehicles.

\subsection{Maneuver Sharing and Coordination Service} 
MSCS is a service allowing connected (and automated) vehicles to share maneuvers. Unlike local planning, MSCS optimizes the planned trajectory by considering the planned trajectory of other vehicles. To achieve this goal, MSCS relies on two communication protocols: one for regular users (e.g., cars and trucks) and one for special vehicles (e.g., ambulances and police cars). The protocol for regular users has a request and response design. The requester will send a MSCM to the maneuver participants detailing all the maneuvers to be performed by each participant. Accordingly, each participant will send a response (a MSCM) to the request (agree or disagree). A single negative response ends the maneuver negotiation (protocol session). A unanimous positive answer leads to the start of the maneuver. For special vehicles, there is no maneuver negotiation. During the maneuver, participants may send a MSCM to cancel the ongoing maneuver (e.g., due to a flat tire). Lastly, a maneuver is considered complete as soon as each participant acknowledges (via a MSCM) the completion of its assigned maneuver.

MSCS allows the requester to send its maneuver request via unicast, groupcast, or broadcast mode. In unicast mode, the requester will negotiate the maneuver with a single vehicle. In groupcast mode, the requester can adjust the signal strength and orientate the signal beam to negotiate a maneuver with a subset of surrounding vehicles. Finally, in broadcast mode, the requester will negotiate a maneuver with all the vehicles within communication range.

\subsection{Maneuver Sharing and Coordination Messages} 

\begin{figure*}
    \centering
    \includegraphics[width=\textwidth]{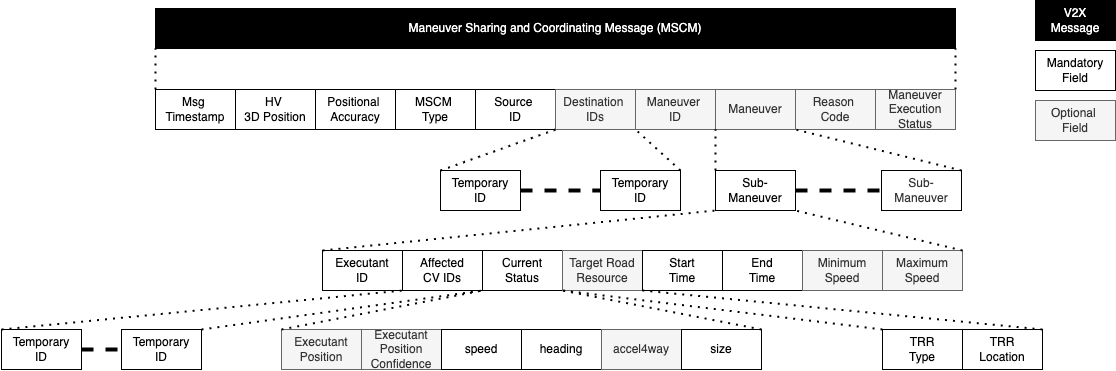}
    \caption{Message Format}
    \label{fig:MSCM}
\end{figure*}

The MSCM consists of an ITS Protocol Data Unit header and containers to include information about the transmitting station (vehicle or infrastructure), vehicles involved in the maneuver negotiation, and maneuvers.

Figure~\ref{fig:MSCM} shows the full structure of a MSCM. The white fields are mandatory; greyed fields are situational (optional), depending on the current stage of the maneuver negotiation (as indicated via the \textit{MSCM Type}). For instance, a MSCM for requesting a maneuver contains fields describing the maneuver (\textit{Maneuver}) and its participants. On the other hand, a MSCM for responding to a request will not contain fields describing the maneuver but fields answering this request (\textit{Reason Code}). 

The \textit{Destination IDs} are identifiers of all the vehicles involved in the maneuver negotiation. Note, \textit{Destination IDs} include the identifier of vehicles requested to perform maneuvers (\textit{Executant IDs}) and of vehicles that do not perform maneuvers (e.g., spectators).

The \textit{Maneuver ID} is the session's identifier used during the maneuver negotiation. A vehicle may be simultaneously involved in multiple maneuver sessions. Thus, the \textit{Maneuver ID} helps to identify which session the vehicle is answering.

The \textit{Maneuver Execution Status} contains an integer describing the status (canceled or completed) of the approved and ongoing maneuver performed by the MSCM transmitter. 

\textit{Maneuver} describes each maneuver (\textit{Sub-Maneuver}) performed by an executant. For instance, a maneuver to overtake requires a first executant to move to a new lane (first \textit{Sub-Maneuver}) to let a second executant overtake the first executant (second \textit{Sub-Maneuver}). Each \textit{Sub-Maneuver} contains information related to its executant (e.g., \textit{Current Status}) and its description (\textit{Target Road Resource}, also known as TRR).

\subsection{V2X Applications and Security}

\subsubsection{Applications}
V2X applications rely on V2X messages as input to warn the driver or to control the vehicle dynamics to avoid road hazards or improve gas consumption. Several safety critical ADAS applications would benefit from using MSCM such as Cooperative Automated Overtaking (CAO) and Cooperative Automated Parking (COP)~\cite{car2019guidance}.

For example, CAVs performing a CAO will benefit from a MSCS. For instance, CAVs get richer information about the maneuver intent, such as the needed portion of the road, the starting time, and the maneuver duration. Also, the MCSC reduces the computation load in each vehicle. Indeed, each vehicle does not need to estimate the trajectory of each surrounding vehicle using its kinematics state. However, note that these applications are still unspecified from a standard perspective.

\subsubsection{Security} 
The MSCS specification includes security requirements such as MSCM's integrity and transmitter's authenticity. Following the IEEE 1609.2~\cite{IEEE1609.2}, the message's integrity and transmitter's authenticity are ensured by digitally signing every MSCM sent. Receivers use the transmitter's public key in the certificate to verify the digital signature attached to the MSCM. 

\section{Attacker Model}
\label{sec:attacker-model}
To facilitate the threat assessment, we formalize the attacker model following the classification proposed in~\cite{monteuuis2018attacker}.

\textit{Internal versus External:} The internal attacker is an authenticated network member that can communicate with other members. The external attacker cannot properly sign her messages, which limits the diversity of attacks. Nevertheless, she can eavesdrop on the V2X broadcast communication.

\textit{Malicious versus Rational:} A malicious attacker seeks no personal benefits from the attacks and aims to harm the members or the functionality of the network. Hence, she may employ any means disregarding corresponding costs and consequences. On the contrary, a rational attacker seeks personal profit and is more predictable in terms of attack means and target.

\textit{Active versus Passive:} An active attacker can generate packets or signals to perform the attack, whereas a passive attacker only eavesdrops on the communication channel (i.e., wireless or in-vehicle wired network).

\textit{Local versus Extended:} An attacker can be limited in scope, even if she controls entities at an intersection (vehicles or base stations), which makes her local. An extended attacker controls scattered entities across the network, thus extending her scope.

\textit{Direct versus Indirect:} A direct attacker reaches its primary target directly, whereas an indirect attacker reaches its primary target through secondary targets. For instance, an indirect attacker may compromise an MSCM through a sensor attack via the planning stack.

\begin{figure}[!b]
    \centering
    \includegraphics[width=\columnwidth]{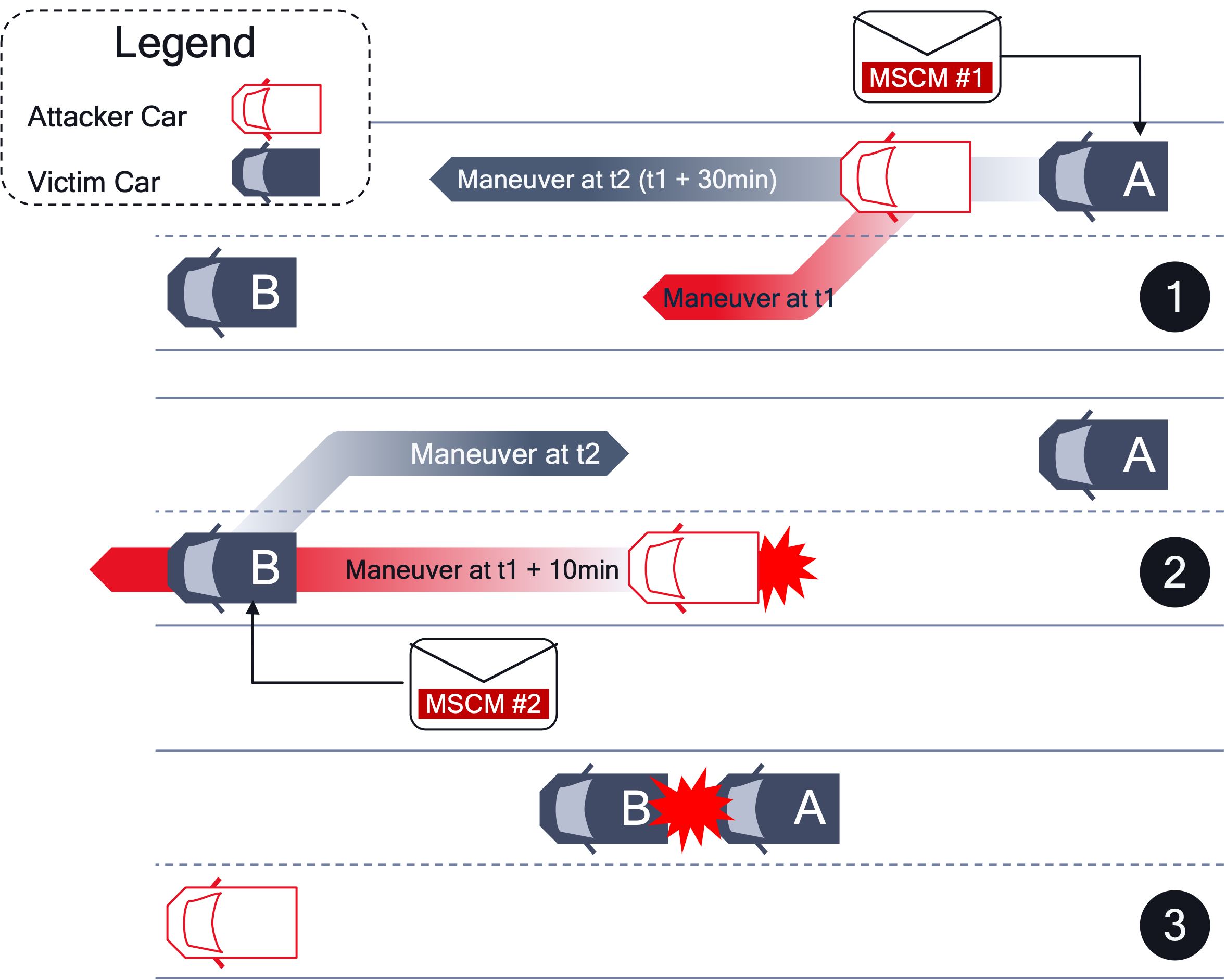}
    \caption{Attack on CAO via MSCS to provoke a cars collision}
    \label{fig:MSCS_Attack}
\end{figure}

Figure~\ref{fig:MSCS_Attack} shows an example of an attack on the MSCS. This example assumes an attacker (white vehicle) proposes overlapping maneuvers to two different vehicles. For each vehicle, the attacker will use a unicast model. For example, in a first maneuver negotiation (time t1), the attacker proposes that vehicle A completes a maneuver at t3. Then, in a second maneuver negotiation (t2), the attacker proposes that vehicle B completes a maneuver at t3. As a result, vehicles A and B are driving towards the same location, resulting in increased safety risk (car collisions) and decreased traffic efficiency. This example demonstrates the importance of assessing data trustworthiness and detecting attacks in MSCS.

\section{Threat Assessment}
\label{sec:threat-assessment}

\begin{table}
\centering
\caption{Risk ratings and criteria~\cite{hadded2020security}}
\begin{tabular}{|l|P|P|P|}
\hline
\multicolumn{1}{|c|}{Criteria}
    & \multicolumn{1}{|c|}{High} 
        & \multicolumn{1}{|c|}{Medium} 
            & \multicolumn{1}{|c|}{Low} \\
\hline
Reproducibility
    & The attack is easily reproducible
        & The attack is reproducible with some limitations
            & The attack is hard to reproduce due to its complexity or operational cost. \\
\hline 
Impact
    & The attack infects the system and can lead to catastrophic damage (e.g., an accident)
        & The attack infects the system and can lead to moderate damage (e.g., traffic jam)
            & The attack has no impacts on the system but can inflict minor harm \\
\hline 
Stealthiness
    & Unknown attack occurs in certain applications
        & The attack needs several misbehavior detectors, message types, or data sources to be detected
            & Broadcasted information readily explain the misbehavior\\
\hline  
\end{tabular}
\label{tab: risk matrix}
\end{table}

\subsection{Methodology}
Several methodologies exist to assess the risk level of an attack. For example, attack trees were used to formalize attacks on V2V communication~\cite{monteuuis2018sara}. However, in our context, the large number of attacks makes the trees too large and unwieldy. Therefore, our methodology follows a matrix approach based on three criteria: \textit{reproducibility}, \textit{impact}, and \textit{stealthiness} (see Table~\ref{tab: risk matrix}). The attack \textit{reproducibility} aims to assess the ease of replicating the attack. Then, the impact measures the impact of the attack on the victim’s car and its surrounding vehicles (i.e., criticality and scalability). Lastly, the attack \textit{stealthiness} assesses the ease by which a driver or a system can detect it. Accordingly, we assess the overall risk level for each threat based on the majority rating among the criteria. For attacks with all three (High, Medium, Low) ratings in the criteria, the overall rating is taken as Medium.

\subsection{Results}
We performed a threat assessment of SAE J3186~\cite{SAEJ3186}, identifying 16 attacks (see Table~\ref{tab:full_use_cases_tara}). As a result, we found eight high, one medium, and seven low-risk attacks.
%IMPORTANT%
% Use the following for camera ready
Although there are more medium-risk and low-risk attacks, some attacks are very easily reproducible, and some have the capability of a very high impact on the MSCS. Hence, we selected a subset of such attacks and presented our findings in Table~\ref{tab:selected_use_cases_tara}. %\textcolor{blue}{The full TARA is available here} \textcolor{red}{change reference~\cite{full_tara}}.

As described in Section~\ref{sec:attacker-model}, the attacker model considered can modify all of the MSCM's fields with any desired value. 

For example, one attack considers an attacker performing the maneuver request and response to create fake maneuvers using multiple pseudonym IDs (spoofing). This information is false, but a receiver cannot corroborate such information without other data sources. For instance, the victim could use its camera to check if the maneuvering vehicle exists.

One attack is an attacker denying all the requests for maneuver (Denial-of-Service attack). The transmitting vehicle could only corroborate against this attack by observing the same behavior numerous times within a long time window or towards a specific transmitting vehicle. 

One attack on \textit{the maximum speed} is when an attacker sets a speed value greatly above the speed limit. However, some special vehicles (e.g., ambulances) sometimes must maneuver above the speed limit. This example shows the complexity of designing robust MBDS for MSCMs.

\subsection{Takeaways}
Most attacks have high \textit{reproducibility} (only one has a medium rating) since they do not require special hardware to perform the attack. in Table~\ref{tab:selected_use_cases_tara}, 3 out of 6 chosen attacks, have high \textit{impact} rating since they have the potential to threaten the lives of drivers and pedestrians. Lastly, these attacks are rated \textit{low} for \textit{stealthiness} as the attacker would be exposing its certificate in the malicious messages and can be easily detected if the suggested defenses are deployed.

Although the attacks we developed have high \textit{reproducibility} and \textit{impact}, we have suggested defense mechanisms that should be able to detect such attacks and help report the malicious actors. However, these defense mechanisms mainly require redundant V2X information from other honest actors surrounding the target vehicle or sensors on the target vehicle. Thus, the defense mechanisms can only be practically applied if the standards allow room for redundant information.

%%%%%%%%%%%%%%%%%%%%%%%%%% DISCUSSION %%%%%%%%%%%%%%%%%%%%%%%%%
\section{Discussion}
\label{sec:discussion}
In this section, we propose standard-related directions to address security gaps identified by the threat assessment.

\subsection{Misbehavior detectors and reporting}
ETSI TR 103 460 and TS 103 759 list a set of misbehavior detectors for BSM. Currently, the TS draft does not specify detectors for the MSCM, leaving that for a future version. However, we can assume that detectors (designed for BSM) also apply to MSCM. For instance, in TR 103 460, the detector, named \textit{implausible speed}, will be the same for both MSCM and BSM.

Additional detectors specific to MSCM will be needed, however. For example, a detector could check if a \textit{Maneuver} contains overlapping \textit{Sub-maneuvers} to prevent car collisions. For instance, an attacker sends a maneuver request with two overlapping maneuvers. The attacker aims to force one vehicle to collide with a second vehicle. At a protocol level, a second detector could check if a maneuver participant keeps declining consecutive maneuver requests within a short time window (e.g., 5 seconds) or sent by a specific requester.

After being detected, a misbehavior report (MBR) may be generated and sent to authorities for further investigation. The ASN.1 definition specified in TS 103 759 is flexible enough to allow for MSCM detectors.

\subsection{Use of MSCM as data source for V2X MBD (and vice versa)}
It can be tempting to use MSCM as data source to detect malicious BSMs (or to use BSMs to detect malicious MSCM). For instance, a CAV reported may have sent BSM data inconsistent with the corresponding MSCM. In detail, the expected maneuver described in the MSCM is inconsistent with the ongoing maneuver depicted by the BSMs. Another example is the inconsistency between the vehicle dimension in the MSCM and the vehicle dimension in the BSM. 

Additionally, using sensors or the mapping stack to detect malicious MSCM will be beneficial. For instance, an attacker may send a MSCM to perform a maneuver on a lane that does not exist. A CAV could detect this attack by looking at the number of lanes in the mapping stack or via the lane detection algorithm performed by its camera. To further improve the MSCS' trustworthiness and prevent attacks on \textit{TRR Location}, extending the IEEE 1609.2 certificate format could be useful to include ego-vehicle capabilities. This extension would allow for (authenticated) sensing and mapping capabilities attestation.

The specification of these detectors, looking at inconsistencies between different message types, will be included in a future version of TS 103 759.

\subsection{Adversarial defense for local planning}
The V2X module of a CAV assumes trustworthy sensor data. This assumption is invalid, considering recent attacks on trajectory prediction~\cite{man2022evaluating}. Indeed, if an external attacker fools the planning stack, a CAV could not trust the maneuvers contained in a MSCM. Also, a CAV could not perform a consistency check between the maneuvers in a received MSCM and the predicted maneuvers from its planning stack.

Recent research proposed defenses. For instance, researchers used adversarial training to increase the robustness of trajectory prediction against adversarial examples~\cite{jiao2022semi}. Mentioning the use of defenses for local planning in standards will decrease the risk of malicious MSCM. Such standardization effort could happen in the ISO TC22 SC32 committee as part of the future ISO 5083.

\section{Conclusion}
\label{sec:conclusion}
Maneuver Sharing and Coordination Service (MSCS) offers to V2X-equipped vehicles the ability to exchange richer data to improve their telematics awareness and safety. However, the security of MSCS and its underlying message set is critical to guarantee quality data. Standardization efforts of MSCS and V2X misbehavior detection and reporting (separately) are ongoing worldwide, but misbehavior protection in MSCS still has to be addressed. In this paper, we summarized a threat assessment done on SAE J3186, which identified 16 attacks with mainly low and high risk levels. Thanks to this assessment, we proposed four directions to consider in ongoing standardization efforts. We hope this work could serve as a starting point to tackle the question of MSCS security by standard organizations and regulators.

% \begin{landscape}
\begin{table*}
\centering
\caption{Threat analysis of selected use-cases from SAE J3186}
\label{tab:selected_use_cases_tara}
\begin{tabularx}{\linewidth}{|p{3.5cm}|p{3cm}|p{3cm}|X|}
\hline
\multicolumn{1}{|c}{Use Case}
    & \multicolumn{1}{|c}{Attacks} 
        & \multicolumn{1}{|c}{Defense}  
            & \multicolumn{1}{|c|}{Risk} \\
\hline
The attacker generates a MSCM with an incorrect ASN.1 format (the message is not decodable).  
	& Omit a mandatory field in the MSCM
	    & \multicolumn{1}{c|}{-}
	        &  \textbf{Overall: \textit{Low}.}
	        \begin{itemize}
    	        \item \textbf{(\textit{High}) Reproducibility:} A malicious transmitter generates MSCMs with an incorrect format.
    	        \item \textbf{(\textit{Low}) Impact :} The message is not decodable. However, the attacker occupies the channel.
    	        \item \textbf{(\textit{Low}) Stealthiness:} An Attacker has to transmit a signed MSCM and hence will be reported and revoked eventually.
	        \end{itemize} \\
\cline{2-4} 
	& Insert a \textit{TRR Location} format that does not match the value contained in the field \textit{TRR Type}
	    & Detect inconsistency between \textit{TRR Type} and \textit{TRR Location}
	        &  \textbf{Overall: \textit{Low}.}
	        \begin{itemize}
    	        \item \textbf{(\textit{High}) Reproducibility:} A malicious transmitter generates MSCMs with an incorrect format.
    	        \item \textbf{(\textit{Low}) Impact:} The message is not decodable. However, the attacker occupies the channel
    	        \item \textbf{(\textit{Low}) Stealthiness:} An attacker has to transmit a signed MSCM and hence will be reported and revoked eventually.
	        \end{itemize} \\
\hline  
The attacker misbehave during the maneuver negotiation (MSCS protocol)
	& The attacker performs the maneuver request and response to create fake maneuvers using pseudonym IDs (spoofing)
	    & Correlate with camera's information
	        &  \textbf{Overall: \textit{High}.} 
	        \begin{itemize}
    	        \item \textbf{\textit{(High)} Reproducibility:} An attacker encodes inaccurate information into the \textit{Maneuver} before signing and transmitting.
    	        \item \textbf{\textit{(High)} Impact:} Surrounding vehicle cannot maneuver if the attacker has planned some (fake) maneuvers
    	        \item \textbf{\textit{(Medium)} Stealthiness:} Onboard sensors should reveal the requester or the responder does not exist at the specified location.
	        \end{itemize} \\
\cline{2-4} 
	& The attacker denies all the request for maneuver
	    & Detect an abnormal number of request for maneuver being denied 
	        &  \textbf{Overall: \textit{High}.} 
	        \begin{itemize}
    	        \item \textbf{\textit{(High)} Reproducibility:} An attacker sets the code to denies some request.
    	        \item \textbf{\textit{(High)} Impact:} Cancelling a maneuver request leads to traffic inefficiency.
    	        \item \textbf{\textit{(Low)} Stealthiness:} An Attacker has to transmit a signed MSCM and hence will be reported and revoked eventually.
	        \end{itemize} \\
\hline  
The attacker inserts an incorrect value in the MSCM-request
	&Set the \textit{maximum speed} with a value way above speed limit (e.g., 200 km/h $>$ 130 km/h)
	    & The \textit{maximum speed} is way above the average speed of surrounding vehicles or the speed limit displayed by the map or perceived by the camera.
	        &  \textbf{Overall: \textit{High}.}
	        \begin{itemize}
    	        \item \textbf{\textit{(High)} Reproducibility:} An attacker inserts a malicious value to the field \textit{maximum speed}
    	        \item \textbf{\textit{(High)} Impact:} Maneuvering vehicles maneuver way above the speed limit (safety risk).
    	        \item \textbf{\textit{(Low)} Stealthiness:} speed value way above the maximal speed limit (implausible value).
	        \end{itemize} \\
\cline{2-4}
	& Attacker request a maneuver on a nonexistent lane by setting an incorrect \textit{LaneOffset} 
	    & Check the number of lanes displayed by the map or perceived by the camera.
	        & \textbf{Overall: \textit{Medium}.}
	        \begin{itemize}
    	        \item \textbf{\textit{(High)} Reproducibility:} An attacker inserts a malicious value to the field \textit{LaneOffset} (located in the container \textit{TRR Location})
    	        \item \textbf{\textit{(Medium)} Impact:} Set the vehicle off the road (safety risk).
    	        \item \textbf{\textit{(Low)} Stealthiness:} An attacker is detectable through its certificate in the MSCM.
	        \end{itemize} \\
\hline
\end{tabularx}
\end{table*}
% \end{landscape}

\begin{table*}
\centering
\caption{Threat analysis of use-cases from SAE J3186}
\label{tab:full_use_cases_tara}
\begin{tabularx}{\linewidth}{|p{1.5cm}|p{3 cm}|p{3 cm}|X|}
\hline
\centering Use Case
    & \multicolumn{1}{|c}{Attacks} 
        & \multicolumn{1}{|c}{Defense}  
            & \multicolumn{1}{|c|}{Risk} \\
\hline
The attacker performs spoofing attacks (ghost vehicle, ghost maneuvers).  
	& Overloads the MSCM with fake \textit{executant ID}s and fake \textit{Sub-Maneuver}s to create a longer processing time for the receivers.
	    & Use the camera to detect ghost vehicle and to check for maneuver consistency
	        &  \textbf{Overall: \textit{High}.}
	        \begin{itemize}
    	        \item \textbf{(\textit{Medium}) Reproducibility:} A malicious transmitter can generate MSCM with an incorrect format.
    	        \item \textbf{(\textit{High}) Impact :} The message is not decodable. However, the attacker occupies the channel
    	        \item \textbf{(\textit{High} Stealthiness:} a fake vehicle with plausible mobility data is hard to detect without the PKI or the use of sensors.
	        \end{itemize} \\
\cline{2-4} 
	& Add overlapping \textit{sub-maneuver}s to provoke a car collision
	    & Check if at least two \textit{sub-maneuver}s are overlapping in time and space
	        &  \textbf{Overall: \textit{High}.}
	        \begin{itemize}
    	        \item \textbf{(\textit{High}) Reproducibility:} Crafting overlapping \textit{sub-maneuver}s does not require advanced knowledge.
    	        \item \textbf{(\textit{High}) Impact:} The message is not decodable. However, the attacker occupies the channel
    	        \item \textbf{(\textit{Low}) Stealthiness:} Attacker has to transmit a signed MSCM and hence will be reported and revoked eventually.
	        \end{itemize} \\
\hline  
The attacker misbehave during the maneuver negotiation (MSCS protocol)
	& The attacker does not answer to some maneuver request
	    & Exclude the attacker and report evidence showing the attacker can transmit V2X message but choose to not participate (e.g., a maneuver request that has been approved by the attacker)
	        &  \textbf{Overall: \textit{High}.} 
	        \begin{itemize}
    	        \item \textbf{\textit{(High)} Reproducibility:} An attacker can encode inaccurate information into the \textit{SensorInformationContainer} before signing and transmitting.
    	        \item \textbf{\textit{(High)} Impact:} Surrounding vehicle cannot maneuver in the absencense of respo
    	        \item \textbf{\textit{(High)} Stealthiness:} Hardly detectable
	        \end{itemize} \\
\hline  
The attacker inserts an incorrect value in the MSCM-request
	&Set the \textit{minimal speed} with a value way below speed limit (e.g., 10 km/h $<$ 130 km/h)
	    & The \textit{minimal speed} is way below the average speed of surrounding vehicles or the speed limit displayed by the map or perceived by the camera.
	        &  \textbf{Overall: \textit{High}.}
	        \begin{itemize}
    	        \item \textbf{\textit{(High)} Reproducibility:} An attacker can insert a malicious value to the field \textit{maximum speed}
    	        \item \textbf{\textit{(High)} Impact:} Maneuvering vehicles will maneuver way below the speed limit (safety risk).
    	        \item \textbf{\textit{(Low)} Stealthiness:} speed value is way below the maximal speed limit (implausible value).
	        \end{itemize} \\
\cline{2-4}
	& Set a static field  (e.g., \textit{executant width}) with a plausible but incorrect value for a single a MSCM to prevent other vehicles to maneuver at a given moment.
	    & \centering Check if the static field value is consistent with other MSCMs or with the BSMs of the attacker
	        & \textbf{Overall: \textit{Low}.}
	        \begin{itemize}
    	        \item \textbf{\textit{(High)} Reproducibility:} Set a fake value in a field does not require advanced knowledge.
    	        \item \textbf{\textit{(Low)} Impact:} Prevent vehicles to maneuver (traffic jam).
    	        \item \textbf{\textit{(Medium)} Stealthiness:} the value is inconsistent with other sources (e.g., MSCM) to detect the origin of the inconsistency 
	        \end{itemize} \\
\cline{2-4}
	& Set the \textit{executant width} with a size much bigger than the lane to prevent other vehicle from maneuvering (e.g., vehicle width  $>$ lane width) 
	    & Check if the vehicle's width is above a threshold  and check for consistency with the width value contained in the BSM sent by the attacker.
	        & \textbf{Overall: \textit{Low}.}
	        \begin{itemize}
    	        \item \textbf{\textit{(High)} Reproducibility:} Set a fake value in a field does not require advanced knowledge.
    	        \item \textbf{\textit{(Low)} Impact:} Prevent vehicles to maneuver (traffic jam).
    	        \item \textbf{\textit{(Low)} Stealthiness:} width value is above the lane width (implausible value)
	        \end{itemize} \\
\cline{2-4}
	& Set the \textit{executant length} with a size much bigger than the length to prevent other vehicle from maneuvering (e.g., vehicle length  $>$ 30m) 
	    & Check if the vehicle's length is above a threshold and check for consistency with the length value contained in the BSM sent by the attacker.
	        & \textbf{Overall: \textit{Low}.}
	        \begin{itemize}
    	        \item \textbf{\textit{(High)} Reproducibility:} Set a fake value in a field does not require advanced knowledge.
    	        \item \textbf{\textit{(Low)} Impact:} Prevent vehicles to maneuver (traffic jam).
    	        \item \textbf{\textit{(Low)} Stealthiness:} length value is an outlier looking at a length distribution for vehicles 
	        \end{itemize} \\
\cline{2-4}
	& Set the maneuver's \textit{starting time} after the \textit{ending time}
	    & Check if \textit{starting time} is set before the \textit{ending time}.
	        & \textbf{Overall: \textit{Low}.}
	        \begin{itemize}
    	        \item \textbf{\textit{(High)} Reproducibility:} Set a fake value in a field does not require advanced knowledge.
    	        \item \textbf{\textit{(Low)} Impact:} The vehicle must process an implausible MSCM data and cannot accept other MSCM (DoS).
    	        \item \textbf{\textit{(Low)} Stealthiness:} the attacker is detectable through its certificate in the MSCM.
	        \end{itemize} \\
\cline{2-4}
	& Set the maneuver's \textit{starting time} before the \textit{Msg Timestamp}
	    & Check if \textit{starting time} is set after the \textit{Msg Timestamp}.
	        & \textbf{Overall: \textit{Low}.}
	        \begin{itemize}
    	        \item \textbf{\textit{(High)} Reproducibility:} Set a fake value in a field does not require advanced knowledge.
    	        \item \textbf{\textit{(Low)} Impact:} The vehicle must process an implausible MSCM data and cannot accept other MSCM (DoS).
    	        \item \textbf{\textit{(Low)} Stealthiness:} the attacker is detectable through its certificate in the MSCM.
	        \end{itemize} \\
\cline{2-4}
	& Set the \textit{Maneuver}'s or the \textit{Sub-Maneuver}'s duration is too long (e.g., \textit{starting time} + the \textit{starting time} $>$ 1 min) preventing other maneuver request
	    & Check if duration between the smallest \textit{Sub-Maneuver}'s \textit{starting time} and the latest \textit{Sub-Maneuver}'s \textit{ending time} is   over a threshold (e.g., 1min0  \textit{starting time} is set after the \textit{Msg Timestamp}.
	        & \textbf{Overall: \textit{High}.}
	        \begin{itemize}
    	        \item \textbf{\textit{(High)} Reproducibility:} Set a fake value in a field does not require advanced knowledge.
    	        \item \textbf{\textit{(High)} Impact:} The attacker prevents other CAVs to perform a maneuver request because the maneuver is still ongoing (DoS).
    	        \item \textbf{\textit{(Low)} Stealthiness:} the attacker is detectable through its certificate in the MSCM.
	        \end{itemize} \\
\hline
\end{tabularx}
\end{table*}

\appendices

\ifCLASSOPTIONcaptionsoff
  \newpage
\fi

\bibliographystyle{IEEEtran}
\bibliography{bib.bib}

\end{document}